\documentclass[preprintnumbers,10pt,nofootinbib]{revtex4}

\usepackage{amsmath,latexsym,amssymb,amsfonts}
\usepackage{graphicx,color}
\usepackage{epstopdf}
\usepackage{bm}
\usepackage{ulem} 
\usepackage{float}
\definecolor{amaranth}{rgb}{0.9, 0.17, 0.31}
\definecolor{purple(munsell)}{rgb}{0.62, 0.0, 0.77}
\definecolor{americanrose}{rgb}{1.0, 0.01, 0.24}
\definecolor{palatinateblue}{rgb}{0.15, 0.23, 0.89}
\definecolor{royalblue(web)}{rgb}{0.25, 0.41, 0.88}
\definecolor{hanpurple}{rgb}{0.32, 0.09, 0.98}
\definecolor{beaublue}{rgb}{0.74, 0.83, 0.9}
\definecolor{carminered}{rgb}{1.0, 0.0, 0.22}
\definecolor{brightpink}{rgb}{1.0, 0.0, 0.5}
\definecolor{vividviolet}{rgb}{0.62, 0.0, 1.0}

\begin{document}

\title{\textbf{A possible origin of the $\alpha$-vacuum as the initial state of the Universe}}

\author{
Pisin Chen$^{a,b,c,d}$\footnote{{\tt pisinchen@phys.ntu.edu.tw}},
Kuan-Nan Lin$^{a,b}$\footnote{{\tt knlinphy@gmail.com}},
Wei-Chen Lin$^{e,f}$\footnote{{\tt archennlin@gmail.com}}
and
Dong-han Yeom$^{a,e,g,h}$\footnote{{\tt innocent.yeom@gmail.com}}
}

\affiliation{
$^{a}$Leung Center for Cosmology and Particle Astrophysics, National Taiwan University, Taipei 10617, Taiwan\\
$^{b}$Department of Physics and Center for Theoretical Sciences, National Taiwan University, Taipei 10617, Taiwan\\
$^{c}$Graduate Institute of Astrophysics, National Taiwan University, Taipei 10617, Taiwan\\
$^{d}$Kavli Institute for Particle Astrophysics and Cosmology,
SLAC National Accelerator Laboratory, Stanford University, Stanford, California 94305, USA\\
$^{e}$Center for Cosmological Constant Problem, Extreme Physics Institute, Pusan National University, Busan 46241, Republic of Korea\\
$^{f}$Department of Physics, Pusan National University, Busan 46241, Republic of Korea\\
$^{g}$Department of Physics Education, Pusan National University, Busan 46241, Republic of Korea\\
$^{h}$Research Center for Dielectric and Advanced Matter Physics, Pusan National University, Busan 46241, Republic of Korea
}

\begin{abstract}
We investigate the cosmological observables using the Euclidean path integral approach. Specifically, we study both the no-boundary compact instantons scenario and the Euclidean wormholes scenario that can induce the creation of two universes from nothing. It is known that perturbations associated with the no-boundary scenario can only be consistent with the Bunch-Davies vacuum. Here we demonstrate that the Euclidean wormholes can allow for a de Sitter invariant vacuum, the so-called $\alpha$-vacuum state, where the Bunch-Davies vacuum is a special case. This therefore provides the $\alpha$-vacuum a geometrical origin. As an aside, we discuss a subtle phase issue when considering the power spectrum related to $\alpha$-vacuum in the closed universe framework.   
\end{abstract}

\maketitle

\newpage

\tableofcontents

\section{Introduction}

While modern cosmology has entered the era of precision cosmology, the inflationary cosmology \cite{Starobinsky:1980te} is currently the most-favored one amongst all existing early universe theories. In an inflationary scenario, quantum fluctuations of the inflaton field were amplified and eventually became the seeds for the large-scale structure formation and the observed cosmic microwave background (CMB) anisotropies. For instance, the latest CMB anisotropy measurement from Planck \cite{Planck:2018vyg} shows a nearly scale-invariant scalar power spectrum with the non-measurements of the primordial non-Gaussianities and of the isocurvature perturbations. These results, together with other measurements, constrain the possible inflationary models and currently are in favor of a simple inflationary model with a Starobinsky-type potential, in which a plateau-like potential generates the quasi-de Sitter phase during the slow-roll period \cite{Martin:2013tda}.

Besides the difference in the actions of the inflaton fields, to compute the theoretical expectations of an inflation model, we also need to assume a specific quantum vacuum state for the scalar perturbations. Since the de Sitter space serves as a good approximation of the exponentially expanding period of inflation, the choice of vacuum is commonly reduced to the discussion of the choice of vacuum in a de Sitter space \cite{Mukhanov:2007zz}. In this sense, the most natural one is to choose the \textit{Bunch-Davies vacuum} \cite{Bunch:1978yq}.

However, interestingly, this is not the unique de Sitter invariant vacuum. If we introduce a kind of Bogoliubov transformation from the Bunch-Davies vacuum, we will obtain a new vacuum that is still de Sitter invariant. This continuous class of vacua is known as the \textit{$\alpha$-vacuum} \cite{Chernikov:1968zm,Mottola:1984ar,Allen:1985ux,Bousso:2001mw}, where $\alpha$ is a parameter that describes a specific vacuum and the Bunch-Davies vacuum is just a special case of the $\alpha$-vacuum.

In inflationary cosmology, the surge of the consideration of $\alpha$-vacuum  was triggered by the discussion of the trans-Planckian effect on the predictions of inflation for cosmological fluctuation spectrum \cite{Brandenberger:1999sw}, where Danielsson suggested that the cutoff of our ignorance of the trans-Planckian physics can be encoded in the choice of initial conditions for the field modes when they start out at cutoff energy scale \cite{Danielsson:2002kx} (see also Ref.~\cite{Shankaranarayanan:2002ax}). As a result, the correction to the cosmological fluctuation spectrum can be of the order of $\mathcal{O}(H/\Lambda)$, where $H$ is the Hubble parameter when the physical momentum of the mode is on a par with the fixed cutoff scale $\Lambda$, which in the most modest case is set to be the Planck scale $M_{Pl}$. While in a pure de Sitter background, the above-mentioned Hubble parameter is a constant for different $k$-modes, in a realistic slow-rolling inflationary model, the slowly changing Hubble parameter induces an oscillatory behavior on the power spectrum, which was first quantitatively estimated in Ref.~\cite{Easther:2002xe}. However, later study shows this modulation on the power spectrum might be too small to be measured \cite{Easther:2004vq}. On the other hand, the $\alpha$-vacuum creates 
distinct features in the non-Gaussianities  \cite{Holman:2007na}, so the existence of the $\alpha$-vacuum might be either supported or strongly limited by future observations once the primordial non-Gaussianities are measured.

Next, one should understand that Danielsson's method is only an effective approach, and the correction to the power spectrum can be smaller depending on different theories considered for quantum gravity. For instance, before the model-independent effective approach to the trans-Planckian effect on the cosmological fluctuation spectrum, studies were carried out in a model-dependent way on the modification of dispersion relations or the non-commutative geometry from string theory. Meanwhile, models from quantum cosmology, for instance, the \textit{no-boundary proposal} (\textbf{NBP}) \cite{Hartle:1983ai} and the \textit{tunneling proposal} \cite{Vilenkin:1982de} were not considered in the discussion of the trans-Planckian effect. 
This is, in fact, not so surprising since in both models, the Bunch Davies vacuum is the only consistent choice of the vacuum state for the matter perturbations \cite{Laflamme:1987mx, Vachaspati:1988as}. Thus, if future observations support the $\alpha$-vacuum, the above two popular models from the quantum cosmology frameworks will be disfavored if no new mechanism generating deviation from the Bunch-Davies vacuum can be introduced in a consistent way. Then, a natural question one can ask is, do there exist models from quantum cosmology that support the $\alpha$-vacuum for the fluctuations?  

We find that the answer to this interesting question is positive by considering the Euclidean wormhole scenario. However, we report that deviation from the Bunch-Davies vacuum is exponentially suppressed in the Euclidean regime, so the imprint on the cosmological observables might be small without fine-tuning. Another result worth mentioning is that since the Euclidean wormhole, like the NBP, is formulated in a closed spacetime, we find a non-trivial phase issue for the mode solutions once we consider the mixing of the positive and negative frequency modes. To our best knowledge, this issue has never been raised since the modulation of the power spectrum by the $\alpha$-vacuum is always carried out in a spatially flat universe framework. 
Only after identifying the correct phase factor, can the power spectrum approach to the spatial flat result in the large $n$ limit. We analytically demonstrate the origin of this $n$-dependent phase issue by using a massless scalar field.

This paper is organized as follows. In Sec.~\ref{sec:HH}, we give a brief introduction to the Euclidean wormhole and summarize our formalism to define and compute the power spectrum of the cosmological perturbations. In Sec.~\ref{sec:res}, we compute the power spectrum and compare the results between that of compact instantons and Euclidean wormholes. The reason for including the $n$-dependent phase factor for the initial condition is discussed. Finally, in Sec.~\ref{sec:con}, we summarize our results and discuss possible future research topics.

\section{\label{sec:HH}Essentials of Euclidean cosmology}

In Euclidean quantum cosmology, one deals with the Euclidean path integral, which is a solution to the Wheeler-DeWitt equation at the formal level. Therefore, in principle, this path integral can describe non-perturbative aspects of the Universe. The Euclidean path integral is the integration between the in-state and the out-state \cite{Hartle:1983ai}. Using this approach, Hartle and Hawking provided a wave function, where there is no dependence on the initial boundary; there exists the final boundary dependence only in the wave function. This proposal of the boundary condition of the Universe is known as the \textit{no-boundary proposal} because there is no initial boundary. Interestingly, according to Halliwell and Hawking \cite{Halliwell:1984eu}, the compact instantons of the no-boundary proposal conform with the \textit{Bunch-Davies state} for density perturbations \cite{Bunch:1978yq}.

This no-boundary proposal is a very elegant and reasonable proposal for the boundary condition of the wave function, but this requires the separation of the Euclidean manifold between the initial and the final hypersurfaces. In other words, the Euclidean manifold for the final state must be regular and compact, such as the poles of the sphere. However, \textit{it is not necessarily true in general} \cite{Chen:2017qeh}. If one considers a kind of excited quantum state \cite{Robles-Perez:2010uvf} or non-trivial dynamics of the matter field \cite{Chen:2017qeh}, the regular and compact boundary conditions of the instantons are not allowed. In a cosmological context, the next candidate instanton, which is regular but non-compact, is the \textit{Euclidean wormholes} that connect from the initial boundary to the final boundary \cite{Hawking:1988ae}. There might be several ways to interpret this Euclidean wormhole; either tunneling from the initial boundary to the final boundary or \textit{tunneling of two universes from nothing} \cite{Chen:2016ask}. In this paper, we will mainly focus on the latter interpretation.

\subsection{Euclidean path integral approach}

The Euclidean path integral is a solution to the Wheeler-DeWitt equation at the formal level. Therefore, in principle, this path integral can describe non-perturbative aspects of the Universe. The Euclidean path integral is the integration between the in-state and the out-state as follows \cite{Hartle:1983ai}:
\begin{eqnarray}
\Psi\left[h_{\mathrm{f}}, \phi_{\mathrm{f}}; h_{\mathrm{i}}, \phi_{\mathrm{i}}\right] = \int \mathcal{D}g_{\mu\nu} \mathcal{D}\phi \; e^{- S_{\mathrm{E}} \left[ g_{\mu\nu}, \phi \right]},
\end{eqnarray}
where $h_{\mathrm{i,f}}$ are the 3-metrics at the initial and the hypersurfaces, $\phi_{\mathrm{i,f}}$ are the field values of the matter field $\phi$ at the initial and the final hypersurfaces, respectively, and $S_{\mathrm{E}}$ is the Euclidean action of the metric $g_{\mu\nu}$ and the matter field $\phi$. This integration is to sum over all geometries and field configurations that connect from $| h_{\mathrm{i}}, \phi_{\mathrm{i}} \rangle$ to $| h_{\mathrm{f}}, \phi_{\mathrm{f}} \rangle$.

From this formal wave function, one can use the steepest-descent approximation for practical computations. One can approximate this integration as a sum of on-shell solutions, or so-called \textit{instantons}:
\begin{eqnarray}
\Psi\left[h_{\mathrm{f}}, \phi_{\mathrm{f}}; h_{\mathrm{i}}, \phi_{\mathrm{i}}\right] \simeq \sum e^{- S^{\mathrm{instanton}}_{\mathrm{E}}}.
\end{eqnarray}
In several examples, the instanton of the in-state $| h_{\mathrm{i}}, \phi_{\mathrm{i}} \rangle$ and the instanton of the out-state $| h_{\mathrm{f}}, \phi_{\mathrm{f}} \rangle$ can be independent of each other. In this case, one can present this as follows:
\begin{eqnarray}
\Psi\left[h_{\mathrm{f}}, \phi_{\mathrm{f}}; h_{\mathrm{i}}, \phi_{\mathrm{i}}\right] \simeq e^{- \left[ S^{\mathrm{instanton}}_{\mathrm{E}} (\mathrm{final}) - S^{\mathrm{instanton}}_{\mathrm{E}} (\mathrm{initial})\right]},
\end{eqnarray}
where $S^{\mathrm{instanton}}_{\mathrm{E}} (\mathrm{initial})$ and $S^{\mathrm{instanton}}_{\mathrm{E}} (\mathrm{final})$ are the Euclidean actions for the initial and final instantons, respectively. If this kind of separation of instantons between the initial and the final states is possible, it is not surprising to write the wave function \textit{only} for the final state:
\begin{eqnarray}
\Psi\left[h_{\mathrm{f}}, \phi_{\mathrm{f}} \right] = \int \mathcal{D}g_{\mu\nu} \mathcal{D}\phi \; e^{- S_{\mathrm{E}} \left[ g_{\mu\nu}, \phi \right]} \simeq e^{- S^{\mathrm{instanton}}_{\mathrm{E}} (\mathrm{final})}.
\end{eqnarray}
In this case, there is no dependence on the initial boundary; there exists the final boundary dependence only in the wave function. This proposal of the boundary condition of the Universe is known as the \textit{no-boundary proposal} because there is no initial boundary.

On the other hand, if we do not exclude the initial boundary, we need to see the generic Euclidean path integral in the following form:
\begin{eqnarray}
\Psi \left[ h_{\mu\nu}^{(2)}, \phi_{0}^{(2)} ; h_{\mu\nu}^{(1)}, \phi_{0}^{(1)} \right] = \int \mathcal{D}g_{\mu\nu} \mathcal{D}\mathcal{\phi} \; e^{-S_{\mathrm{E}}[g_{\mu\nu},\phi]},
\end{eqnarray}
where $g_{\mu\nu}$ is the metric, $\phi$ is a matter field, and we sum over all regular Euclidean geometries that have boundary values $\partial g_{\mu\nu} = h_{\mu\nu}^{(1)} \cup h_{\mu\nu}^{(2)}$ and $\partial \phi = \phi_{0}^{(1)} \cup \phi_{0}^{(2)}$.

\begin{figure}[h]
\begin{center}
\includegraphics[scale=0.4]{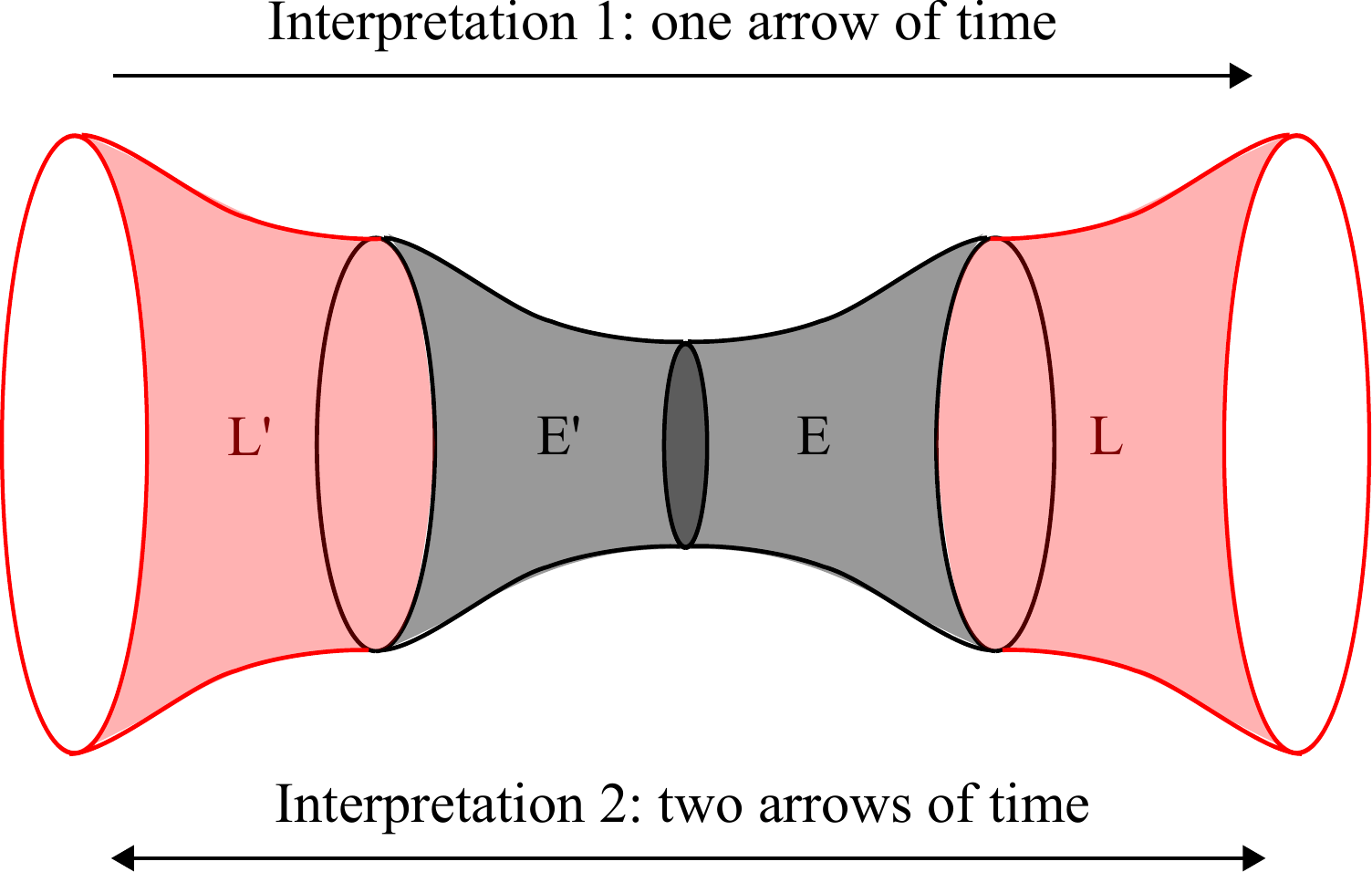}
\caption{\label{fig:interp}Interpretation of Euclidean solutions with two boundaries. $\mathrm{E} \cup \mathrm{E}'$ corresponds the Euclidean manifold that connects two Lorentzian manifolds $\mathrm{L}$ and $\mathrm{L}'$. One can interpret either there exists one arrow or two arrows of time.}
\end{center}
\end{figure}

Since this path integral has two boundaries, there can be two interpretations (Fig.~\ref{fig:interp}):
\begin{itemize}
\item[--] 1. \textit{One} arrow of time: A universe with boundary values $\partial g_{\mu\nu} = h_{\mu\nu}^{(1)}$, $\partial \phi = \phi_{0}^{(1)}$ tunnels to boundary values $\partial g_{\mu\nu} = h_{\mu\nu}^{(2)}$, $\partial \phi = \phi_{0}^{(2)}$, or vice versa.
\item[--] 2. \textit{Two} arrows of time: Two universes with boundary values $\partial g_{\mu\nu} = h_{\mu\nu}^{(1,2)}$ and $\partial \phi = \phi_{0}^{(1,2)}$ are created from nothing.
\end{itemize}
In this paper, we will follow the second interpretation \cite{Chen:2016ask,Chen:2017qeh}.

\subsection{Background solutions}

As a simple model, let us consider a Euclidean action consisting of the standard Einstein-Hilbert term and the kinetic and potential energies of the inflaton:
\begin{eqnarray}
S_{\mathrm{E}} = \int \sqrt{+g} d^{4}x \left[ -\frac{R}{16\pi} + \frac{1}{2} \left( \partial_{\mu} \Phi \right)^{2} + U(\Phi) \right],
\end{eqnarray}
where $\Phi$ is the scalar inflaton with the potential $U(\Phi)$. In addition, we assume the background metric to be a homogeneous, isotropic, and closed universe:
\begin{eqnarray}
ds_{\mathrm{E}}^{2} = \sigma^{2} \left( d\tau^{2} + a^{2}(\tau) d\Omega_{3}^{2} \right),
\end{eqnarray}
where $d\Omega_{3}^{2} = d\chi^{2} + \sin^{2}\chi ( d\theta^{2} + \sin^{2} \theta d\varphi^{2} )$ (with angle variables $\chi$, $\theta$, and $\varphi$) and $\sigma^{2} = 8\pi U_{0}/3$ with a constant $U_{0}$.

After redefining $\phi = \sqrt{4\pi/3} \Phi$ and $V = U/U_{0}$, one can derive the equations of motion for the background geometry and the inflaton:
\begin{eqnarray}
\dot{a}^{2} - 1 + a^{2} \left( - \dot{\phi}^{2} + V\left(\phi\right) \right) &=& 0,\label{bgeomE1}\\
\ddot{a} + 2 a \dot{\phi}^{2} + a V &=& 0,\label{bgeomE2}\\
\ddot{\phi} + 3 \frac{\dot{a}}{a} \dot{\phi} - \frac{1}{2} \frac{dV}{d\phi} &=& 0,\label{bgeomE3}
\end{eqnarray}
where dots denote differentiation for the Euclidean time $\tau$.

\paragraph{Compact instantons} In the no-boundary proposal, one requires the regular and compact condition of the manifold. So, for example, one can present the compactness condition such that $a(0) = 0$. Due to the regularity of the solution, this requires $\dot{a}(0) = 1$ and $\dot{\phi}(0) = 0$.

As a simple demonstration, let us consider $U(\Phi) = \lambda^2 U_{0}$ and $\Phi$ is a constant. Then, the solution becomes
\begin{eqnarray}
a = \lambda^{-1}\sin (\lambda\tau).
\end{eqnarray}
The location for the Wick-rotation is at $\tau = \pi/2\lambda$. After the Wick-rotation to the Lorentzian time $\tau \rightarrow \pi/2\lambda + it$, the result becomes
\begin{eqnarray}
a = \lambda^{-1}\cosh (\lambda t).
\end{eqnarray}

\paragraph{Euclidean wormholes} In Euclidean gravity, there are several models of Euclidean wormholes \cite{Chen:2016ask,Chen:2017qeh}.
\begin{itemize}
\item[(1)] \textit{Conformal gravity}: From the Wheeler-DeWitt equation of Einstein gravity with a conformally coupled scalar field, one can introduce the separation of variables for the entire wave function \cite{Hartle:1983ai,Robles-Perez:2010uvf}. Thanks to this separation of variables, one can specify the energy eigenvalues $E_n = n+1/2$, where $n = 0, 1, 2, 3, ...$ and $a$ should satisfy
\begin{eqnarray}
    \dot{a}^2 = 1-\frac{\Lambda}{3} a^2-\frac{2E_n}{a^2},
\end{eqnarray}
where $\Lambda$ is the cosmological constant. The on-shell solution of this equation has two turning points: $a_{\mathrm{min}}^2\simeq 2E_n$ and $a_{\mathrm{max}}^2\simeq 3/\Lambda$, where the former corresponds to the throat of the Euclidean wormhole and the latter determines the Wick-rotation point.

\item[(2)] \textit{String-inspired theory}: Euclidean wormholes are also possible in the axion-induced model \cite{Giddings:1987cg} given by the Euclidean action
\begin{eqnarray}
    S_{E}=\int d^4x\sqrt{+g}\left(-\frac{R}{16\pi}+\frac{1}{2}G_{IJ}(\phi)\nabla^{\mu}\phi^{I}\nabla_{\mu}\phi^{J}\right),
\end{eqnarray}
where $I,J$ denotes different species of scalar fields and $G_{IJ}$ can have negative signs in Euclidean signatures. Thanks to the negative-sign contributions, $a$ should satisfy
\begin{eqnarray}
    \dot{a}^2 = 1-\frac{C}{a^4},
\end{eqnarray}
where $C$ is a positive constant. This allows a turning point at $a_{\mathrm{min}}^2 = \sqrt{C}$, which corresponds to the throat of an asymptotically flat Euclidean wormhole. Further investigations of Euclidean wormholes in the anti-de Sitter space are in \cite{Maldacena:2004rf,Arkani-Hamed:2007cpn}.

\item[(3)] \textit{Kinetic effects of a scalar field}: For simplicity, let us consider a free scalar field with $U(\Phi) = \lambda^2U_{0}$ case, where $\lambda$ is a rescaled cosmological constant. Then, the scalar field satisfies
\begin{eqnarray}
\frac{d\phi}{d\tau} = i \frac{A}{a^{3}},
\end{eqnarray}
where $A$ is a constant. Hence,
\begin{equation}
\dot{a}^2 = 1 - \lambda^2a^2 - \frac{A^2}{a^4}
\end{equation}
should be satisfied in the Euclidean domain. This allows two turning points: $a_{\mathrm{min}}^2 \simeq |A|$ and $a_{\mathrm{max}}^2 \simeq \lambda^{-2}$ for real-valued $A$ \cite{Chen:2016ask,Chen:2017qeh}. The former corresponds to the throat of an asymptotically de Sitter Euclidean wormhole and the latter is the Wick-rotation point. Note that this kind of Euclidean wormhole can be generalized to the models with a non-trivial potential with inflationary scenario \cite{Chen:2019cmw} or even in the anti-de Sitter background \cite{Kang:2017jmq}.
\end{itemize}

\begin{figure}[tbp]
\centering 
\includegraphics[width=.45\textwidth]{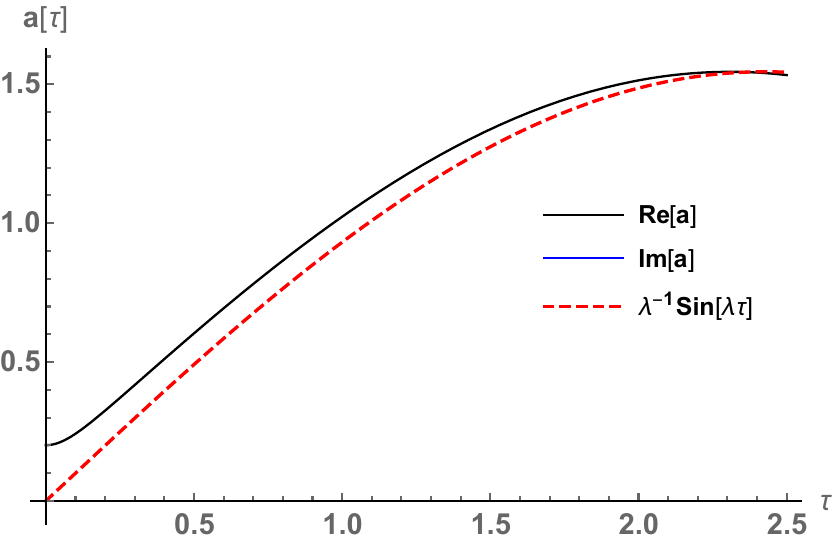}
%\hfill
%\includegraphics[width=.45\textwidth]{phiE.pdf}
\caption{\label{EWH-bg-a}The black curve is the real part and the blue curve is the imaginary part of $a(\tau)$, where we assume an Euclidean wormhole solution following the model in \cite{Chen:2016ask,Chen:2017qeh} with $a_{\mathrm{min}}\simeq 0.200854$ and $a_{\mathrm{max}}\simeq 1.54488$, while the red dashed curve is $a(\tau)$ of a compact instanton with the same $a_{\mathrm{max}}\simeq 1.5451$.}
\end{figure}

In summary, Euclidean wormholes can be described by the following equation of motion:
\begin{equation}
    \dot{a}^2 = 1 - B a^2 - \frac{C}{a^n},
\end{equation}
where $B$ and $C$ are positive constants and $n$ is a positive number. Here, $B$, $C$, and $n$ depend on a specific model. In this paper, we will focus on the case that $n = 4$, but our formalism does not depend on a specific model of the Euclidean wormhole. $C$ determines the throat size of the wormhole $a_{\mathrm{min}}^2 \simeq C^{2/n}$ and $B$ determines the Wick-rotation point $a_{\mathrm{max}}^2 \simeq 1/B$. Note that $C=0$ provides a compact instanton according to the no-boundary proposal. Fig.~\ref{EWH-bg-a} shows an example of the Euclidean wormhole solution as well as the compact instanton following the model in \cite{Chen:2016ask,Chen:2017qeh}. Here, one can notice that the Euclidean wormhole is specified by the model parameters $a_{\mathrm{min}}$ and $a_{\mathrm{max}}$.

\subsection{Perturbations}

In cosmology, in addition to the background geometry and the inflaton field, cosmological perturbations are also important since they are essential for the structure formation. In this section, we introduce the perturbations to the minisuperspace metric following Halliwell-Hawking prescription \cite{Halliwell:1984eu}:
\begin{eqnarray}
ds^{2} = \sigma^{2} \left[ - \left( N^{2} - N_{i} N^{i} \right) dz^{2} + 2 N_{i} dx^{i} dz + h_{ij} dx^{i}dx^{j} \right],
\end{eqnarray}
where
\begin{eqnarray}
N &=& N_{0} \left[ 1 + \frac{1}{\sqrt{6}} \sum_{nlm}g_{nlm}Q_{nlm} \right],\\
N_{i} &=& a \sum_{nlm}\left[ \frac{1}{\sqrt{6}} k_{nlm} \left(P_{i}\right)_{nlm} + \sqrt{2} j^{o}_{nlm} \left( S^{o}_{i} \right)_{nlm} + \sqrt{2} j^{e}_{nlm} \left( S^{e}_{i} \right)_{nlm} \right],\\
h_{ij} &=& a^{2} \left[ \gamma_{ij} + \sum_{nlm}\left( \frac{\sqrt{6}}{3} a_{nlm} Q_{nlm} \gamma_{ij} + \sqrt{6} b_{nlm} \left( P_{ij} \right)_{nlm} + \sqrt{2} c^{o}_{nlm} \left( S_{ij}^{o} \right)_{nlm} + \sqrt{2} c^{e}_{nlm} \left( S_{ij}^{e} \right)_{nlm} \right. \right. \nonumber \\
&& \left. \left.+ 2 d^{o}_{nlm} \left( G_{ij}^{o} \right)_{nlm} + 2 d^{e}_{nlm} \left( G_{ij}^{e} \right)_{nlm} \right) \right],\\
\Phi &=& \sqrt{\frac{3}{4\pi}} \phi + \sqrt{\frac{3\pi}{2}} \sum_{nlm} f_{nlm} Q_{nlm},
\end{eqnarray}
and $\gamma_{ij} dx^{i}dx^{j} = d\Omega_{3}^{2}$. Here, $\{a_{n\ell m},b_{n\ell m},c_{n\ell m},d_{n\ell m},g_{n\ell m},k_{n\ell m},j_{n\ell m},f_{n\ell m} \}$ are perturbations beyond the minisuperspace, the superscript $o\;(e)$ denotes odd (even) parity, $Q_{n\ell m}$ is the scalar harmonic on a 3-sphere of the unit radius, $\{(P_{i})_{n\ell m},(S_i)_{n\ell m}\}$ are vector harmonics, and $\{(P_{ij})_{n\ell m},(S_{ij})_{n\ell m},(G_{ij})_{n\ell m}\}$ are tensor harmonics. 

As usual, $z$ is real-valued in the metric. The Euclidean time is defined by $d\tau^2=-N_{0}^2dz^2$ for an imaginary-valued $N_0$, while the Lorentzian time is defined by $dt^2=N_{0}^2dz^2$ for a real-valued $N_0$. Therefore, the conventional Wick-rotation corresponds to taking $d\tau \rightarrow idt$, which casts Eqs.~\eqref{bgeomE1}, \eqref{bgeomE2}, and \eqref{bgeomE3} into the Lorentzian form:
\begin{eqnarray}
{\dot{a}}^{2} + 1 - a^{2} \left( {\dot{\phi}}^{2} + V\left(\phi\right) \right) &=& 0,\\
{\ddot{a}} + 2 a {\dot{\phi}}^{2} - a V &=& 0,\\
{\ddot{\phi}} + 3 \frac{\dot{a}}{a} {\dot{\phi}} + \frac{1}{2} \frac{dV}{d\phi} &=& 0,
\end{eqnarray}
where dots now denote differentiation with respect to the Lorentzian time $t$. In addition, in the slow-roll limit, one can approximate the equation for the matter perturbation as follows \cite{Halliwell:1984eu}:
\begin{eqnarray}
\ddot{f}_{nlm} + 3 \frac{\dot{a}}{a} \dot{f}_{nlm} + \left( \frac{1}{2} \frac{d^{2}V}{d\phi^{2}} + \frac{n^{2} - 1}{a^{2}} \right) f_{nlm} \simeq 0.
\end{eqnarray}

\subsection{Wave function of matter perturbation}

Cosmological observables are related to the expectation values; as in quantum mechanics, these expectation values are determined by the wave functions. In quantum cosmology, the wave function of the universe can be factorized as:
\begin{eqnarray}
    \Psi \left[ a, \phi, \delta \phi \right] = \Psi_{\mathrm{bg}} \left[a, \phi \right] \prod_{nlm} \psi_{nlm} \left[ f_{nlm}; a, \phi \right],
\end{eqnarray}
where $\Psi_{\mathrm{bg}}$ is the wave function of the background fields $a$ and $\phi$, and $\psi_{nlm}$ is the wave function of the matter perturbation ${f}_{nlm}$ with metric determined by the background fields $(a, \phi)$. The matter wave function $\psi_{nlm}$ can be evaluated from the Euclidean path integral:
\begin{eqnarray}
    \psi_{nlm}= \int \mathcal{D}f_{nlm} \text{exp}\left[ -\int^{\tau_{\mathrm{f}}}_{\tau_{\mathrm{i}}} d\tau L_{nlm} \right]
   \simeq A_{nlm} \text{exp}\left[ -\int^{\tau_{\mathrm{f}}}_{\tau_{\mathrm{i}}} d\tau L_{nlm} \right],
    \label{formal matter wave function}
\end{eqnarray}
where $A_{nlm}$ is a normalization factor, $L_{nlm}$ is the Euclidean Lagrangian of the matter perturbation, and the steepest-descent approximation is applied in the last step. In the Euclidean wormhole considerations, we regard that $\tau_{\mathrm{i}}$ corresponds to the wormhole throat ($a_{\mathrm{min}}$) and $\tau_{\mathrm{f}}$ corresponds the endpoint of the integral that can be chosen arbitrarily. We will evaluate the wave function after the horizon crossing time and check whether our result is independent of the time choice or not. Also, our result must be consistent with the standard result in quantum field theory. 
After integrating the Euclidean action, one obtains
\begin{eqnarray}
    \psi_{nlm} \simeq A_{nlm} \exp \left[ \left. -\frac{1}{2} a^3 f_{nlm} \dot{f}_{nlm} \right|^{\tau_{\mathrm{f}}}_{\tau_{\mathrm{i}}} \right]
    = B_{nlm} \exp \left[ - \left.\frac{1}{2} a^{3} f_{nlm} \dot{f}_{nlm}\right|_{\tau_{\mathrm{f}}} \right],
\end{eqnarray}
where $B_{nlm}$ is a normalization constant that absorbs the constant contribution of the Euclidean action at $\tau_{\mathrm{i}}$.

We must regard $\psi_{nlm}$ as a functional of $f_{nlm}$ and hence we need to regard $f_{nlm}$ as an operator $\hat{f}_{nlm}$. Although we know the value of the wave function, within the constraint of the value, there are infinitely many possibilities to choose the functional dependence between $\hat{f}_{nlm}$ and $\psi_{nlm}$. To provide a probability distribution that is consistent with the results of the quantum field theory, one needs to consider the quantum mode expansion:
\begin{eqnarray}
\hat{f}_{nlm} = f_{nlm} \hat{a}_{nlm} + f^{*}_{nlm} \hat{a}^{\dagger}_{nlm},
\end{eqnarray}
where $a_{nlm}$ and $a_{nlm}^{\dagger}$ are annihilation and creation operators of a certain vacuum, respectively. If we impose the condition that the wave function is annihilated by the operator $a$, we obtain the result \cite{Laflamme:1987mx}:
\begin{eqnarray}
\psi_{nlm} \left[ \hat{f}_{nlm} \right] = C_{nlm} \exp \left[ - \left.\left( \frac{a^{3} \dot{f}_{nlm}}{2 f_{nlm}} \right)\right|_{\tau_{\mathrm{f}}} \hat{f}_{nlm}^{2} \right],
\end{eqnarray}
and hence, the wave function has a Gaussian-like form. This is \textit{a kind of} vacua, but this is not guaranteed whether this vacuum is the Bunch-Davies state or not. Indeed, the properties of the vacuum are related to the background manifold. We will investigate this problem later.

Equivalently, if we Wick-rotate the wave function to the Lorentzian time $t$, we present as follows:
\begin{eqnarray}
\psi_{nlm} \left[ \hat{f}_{nlm} \right] = C_{nlm} \exp \left[ i \left.\left( \frac{a^{3} {\dot{f}}_{nlm}}{2 f_{nlm}} \right)\right|_{t_{\mathrm{f}}} \hat{f}_{nlm}^{2} \right].
\end{eqnarray}
By defining $f_{nlm} =v_{nlm}/a$ and $d\eta = dt/a$, the Lorentzian equations of motion of $v_{nlm}(\eta)$ is then
\begin{eqnarray}
    {v}''_{nlm} = - \left[ n^2 - 1 - \frac{a''}{a}+\frac{a^2}{2} \frac{d^{2}V}{d\phi^{2}} \right] v_{nlm},
\end{eqnarray}
where $'$ denotes differentiation with respect to $\eta$.

After some computations (see Appendix), we obtain the following formula:
\begin{eqnarray}\label{PowerSpectrum}
\mathcal{P}(n)=2\pi^2P(n) = n \left( n^{2} - 1 \right) \left\langle \left| \hat{f}_{n} \right|^2 \right\rangle = \frac{n \left( n^{2} - 1 \right)}{ 2a^{2} \mathrm{Re} \left[ -i \frac{v'_{n}}{v_{n}} \right]},
\end{eqnarray}
where $'$ denotes a differentiation with respect to $\eta$. Here, $v_n=af_n$ and $P(n)$ has the conventional normalization.

\section{\label{sec:res}Quantum states with Euclidean geometric origins}

Now we are ready to compute the power spectrum of a scalar field. We will compare our result from the Euclidean path integral approach with the results from the quantum field theory in the flat space limit, and we will further ask the physical meaning as well as their consistency.

\subsection{Reproduction of the Bunch-Davies state from the no-boundary proposal} 

First, let us compute the power spectrum of the compact instantons (no-boundary proposal).

\subsubsection{Computations from instantons}

In the no-boundary proposal, the Euclidean scale factor is $a(\tau)=\lambda^{-1}\sin(\lambda\tau)$, and the perturbation mode $f_{n}(\tau)$ satisfies the equation of motion:
\begin{eqnarray}\label{EoM_mode}
    \ddot{f}_{n}(\tau) = - 3 \frac{\dot{a}}{a}\dot{f}_n +
   \left( \frac{n^2-1}{a^2}+\mu^2 \right) f_n,
\end{eqnarray}
where $\lambda$ is a cosmological constant and $\mu$ is the inflaton mass. After we solve the equation, we obtain the solution:
\begin{eqnarray}\label{LaflammeSol}
    f_n(\tau)= D_n(\zeta-\zeta^2)^{(n-1)/2}{}_{2}F_{1}\big(n-\nu,n+\nu+1;n+1;\zeta\big),
\end{eqnarray}
where
\begin{eqnarray}
    \nu=-\frac{1}{2}+\sqrt{\frac{9}{4}-\frac{\mu^2}{\lambda^2}},
    \quad
    \zeta=\frac{1-\cos(\lambda\tau)}{2},
\end{eqnarray}
$D_n$ is a normalization constant, and ${}_{2}F_{1}$ is the hypergeometric function \cite{Laflamme:1987mx}. This solution satisfies the Euclidean initial conditions: $\dot{f}_{1}(\tau=0)=0$ and $f_{n\geq 2}(\tau=0)=0$, as required by the regularity condition in the no-boundary proposal implied by Eq.~(\ref{EoM_mode}). By performing analytic continuation $\tau\rightarrow \pi/2\lambda+it$, one obtains the corresponding Lorentzian solution. It is known that this solution picks out the Bunch-Davies vacuum state in a closed universe \cite{Halliwell:1989myn}.

In our numerical treatment, we set the Euclidean initial conditions as follows to make the equation of motion regular \cite{Chen:2017aes}:
\begin{eqnarray}\label{EInitialCond}
    f_{n}(\tau_i)=e^{-i\delta_{n}}\frac{1}{2} \epsilon \tau_{i}^2,\quad
    \dot{f}_{n} (\tau_i)=e^{-i\delta_{n}}\epsilon \tau_{i},
\end{eqnarray}
where $\delta_n$ is an arbitrary phase that can be mode-dependent. Here, $\tau_i\ll 1$ and $\epsilon$ is an arbitrary parameter. (In our work, we set\footnote{The reason for this choice will be discussed in the following sections.} $\delta_n=n\pi/2,\epsilon=\tau_i=10^{-3}$.) Fig.~\ref{Pf-NBP-compare} (left) shows that the scalar power spectrum obtained based on our numerical treatment agrees well with that obtained according to the above analytic formula.

\subsubsection{Dependence on the evaluation time}

In general, the power spectrum is both mode-dependent and time-dependent. However, in certain situations, the time dependence only results in a trivial amplitude shift. For example, in a spatially flat Lorentzian universe with power-law $(a=a_0t^{p})$ inflation, the positive frequency Bunch-Davies perturbation mode $U_{k}(\Tilde{\eta})/a(\Tilde{\eta})$, which now has a continuous mode $k$ instead of the discrete one $n$ and the conformal time $\Tilde{\eta}\in(-\infty,0^{-})$, is given by
\begin{eqnarray}
    U_{k}(\Tilde{\eta})=\sqrt{\frac{-\pi\Tilde{\eta}}{4}}i^{\nu+1/2}H_{\nu}^{(1)}(-k\Tilde{\eta}),
\end{eqnarray}
where $H_{\nu}^{(1)}$ is the Hankel function and $\nu=3/2+1/(p-1)$. In this case, the Hubble parameter $H(t)=\Dot{a}(t)/a(t)$ is time-dependent:
\begin{eqnarray}
    H(y)y^{-\frac{\epsilon}{1-\epsilon}}=\frac{1}{\epsilon}\big(\frac{y}{k\epsilon}\big)^{\frac{\epsilon}{1-\epsilon}}y^{-\frac{\epsilon}{1-\epsilon}}=H_{*}=const.,
\end{eqnarray}
where $y=k/aH=-k(1-\epsilon)\Tilde{\eta}$, $\epsilon=1/p$, and $H_{*}$ here is the Hubble parameter evaluated at the horizon-crossing: $y_{*}=k/a(t_{*})H(t_{*})=1$, which is a constant. 

For $\epsilon=0$, which corresponds to the massless de Sitter inflation in a flat universe, $H=H_{*}$. The corresponding scalar power spectrum $P(k)=k^3|U_{k}|^2/(2\pi^2a^2)$ for a given mode $k$ evaluated at any moment in time $y$ is
\begin{eqnarray}
    P(k)=\frac{H^2}{4\pi^2}(1+y^2)=\frac{H_{*}^2}{4\pi^2}(1+y^2).
\end{eqnarray}
Therefore, if we choose the horizon-crossing time, $y=1$, we obtain $P(k)=H_{*}^2/2\pi^2$; if we choose the super-horizon limit, $y \ll 1$, we obtain $P(k)=H_{*}^2/4\pi^2$, which is the standard result for matter perturbations in the Bunch-Davies state of a spatially flat, massless de Sitter universe. In this case, the power spectra at the horizon crossing time and at the super-horizon limit simply differ by a trivial factor of 2 \cite{Kinney:2005vj}.

\subsubsection{Scale-invariance of the power spectrum}

In the no-boundary proposal, the Hubble parameter $H(t)=\lambda\tanh(\lambda t)$ is time-dependent, and, e.g., it has the horizon-crossing value $H(t_{n})=\lambda n/\sqrt{n^2+1}$, which is mode-dependent. By setting $\lambda=H_{*}$, where $H_{*}$ is the horizon-crossing value for the Hubble parameter in a flat universe, $H(t_{n})=H_{*} n/\sqrt{n^2+1}$. Therefore, for large modes $n\gg 1$, the curvature term $``1"$ is negligible, which results in $H(t_{n})\simeq H_{*}$ becomes approximately mode-independent, and this recovers the scale-invariance of the spectrum.

On the other hand, as far as the time evolution of the comoving Hubble radius $(1/aH)$ is concerned, the spatial curvature of the Universe can cause observable differences around the onset of inflation \cite{Bonga:2016iuf}. Since perturbation modes with small $n$ (long wavelength) leave the horizon earlier than modes with large $n$ (short wavelength), these small $n$ modes can carry imprints of the curvature before the curvature effect is diluted away by the rapid expansion of the Universe. As a result, the power spectrum in the no-boundary proposal is suppressed at small $n$ due to the positive spatial curvature, while the spectrum is always scale-invariant in the absence of curvature.

\begin{figure}[tbp]
\centering 
\includegraphics[width=.45\textwidth]{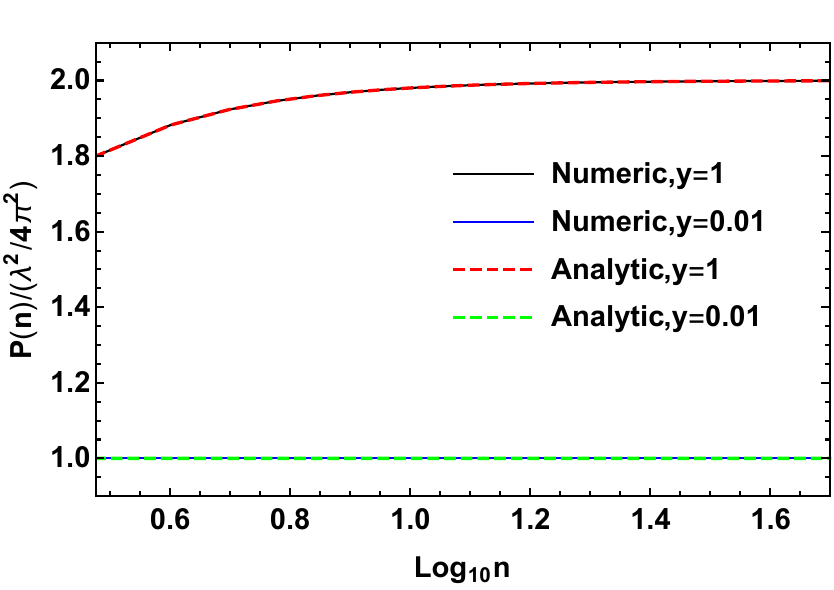}
\hfill
\includegraphics[width=.45\textwidth]{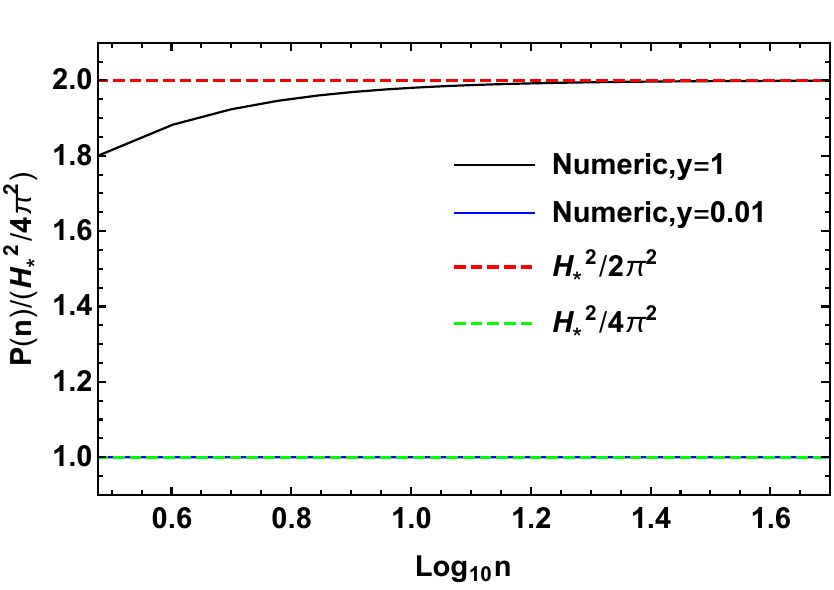}
\caption{\label{Pf-NBP-compare} Left: The scalar power spectrum based on our numerical treatment agrees well with analytic results evaluated at different time $y=n/aH$. Right: In the no-boundary proposal, scale-invariant behavior is recovered at large $n$, while there is suppression at small $n$. The invariant values are indicated by the horizontal dashed lines.
}
\end{figure}

The above discussions of the scalar power spectrum of the no-boundary instantons are summarized in Fig.~\ref{Pf-NBP-compare} (right):
\begin{itemize}
\item[-- 1.] The power spectrum is also time-dependent, where the power spectra at the horizon crossing time and at the super-horizon limit differ by a factor of ``2'' \cite{Kinney:2005vj} in the large $n$ limit.
\item[-- 2.] The scale-invariance is recovered at large $n$.
\item[-- 3.] The power spectrum is suppressed at small $n$ which was observed in \cite{Chen:2017aes,Yeom:2017ikw,Chen:2019mbu}.
\end{itemize}
Now we conclude that our numerical treatment is consistent with the analytic solution as well as various theoretical expectations.

\subsection{Existence of the $\alpha$-vacuum state in Euclidean wormholes}

In the previous subsection, we reported that our numerical computations are consistent with analytic solutions as well as previously known theoretical expectations. Now, using our technique, we will extend the computations to the Euclidean wormholes.

Now one can ask what is the \textit{crucial difference} between the no-boundary proposal and the Euclidean wormholes. As mentioned in the previous subsection, the no-boundary proposal imposes initial conditions on the matter perturbation. However, if we relax the no-boundary proposal, e.g., considering the Euclidean wormhole solution, we will have more freedom to choose the initial conditions which may allow the possibility of going beyond the usual Bunch-Davies state  \cite{Chernikov:1968zm,Mottola:1984ar,Allen:1985ux,Bousso:2001mw}.

Interestingly, in our setup, we already chose operators for a vacuum condition. Therefore, it is very reasonable to conclude that the Euclidean wormholes may explain a kind of vacuum state which goes beyond the Bunch-Davies state. If we further impose the de Sitter invariance condition, we will show that the Euclidean wormholes allow the $\alpha$-vacua for matter perturbations.

\subsubsection{Standard $\alpha$-vacuum in flat de Sitter space}\label{Standard alpha}

Standard discussions of $\alpha$-vacuum involve a mode-independent $\alpha$-parameter with $\mathrm{Re}(\alpha)<0$ and a massive scalar field in a spatially flat inflationary universe. $\alpha$-vacuum is de Sitter invariant and its corresponding mode function is given by the Mottola-Allen transformation \cite{Mottola:1984ar,Allen:1985ux}:
\begin{eqnarray}
    u_{k}\left(\Tilde{\eta}\right) = A_{k}U_{k} \left(\Tilde{\eta}\right)+B_{k}U^{*}_{k}\left(\Tilde{\eta}\right),
\end{eqnarray}
where
\begin{eqnarray}
\left|A_{k}\right|^2-\left|B_{k}\right|^2=1
\end{eqnarray}
is the normalization condition, $\Tilde{\eta}\in(-\infty,0^{-})$,
\begin{eqnarray}
    U_{k}\left(\Tilde{\eta}\right)=\sqrt{\frac{-\pi\Tilde{\eta}}{4}}i^{\nu+1/2}H_{\nu}^{(1)}\left(-k\Tilde{\eta}\right),
\end{eqnarray}
and
\begin{eqnarray}
\nu=\sqrt{\frac{9}{4}-\frac{\mu^2}{H_{*}^2}}.
\end{eqnarray}
Here, $U_k$ is the standard positive frequency Bunch-Davies mode and $(A_k,B_k)=N_{\alpha}(1,e^{\alpha})$ are the Bogoliubov coefficients with $N_{\alpha}=(1-e^{\alpha+\alpha^{*}})^{-1/2}$ \cite{Bousso:2001mw}.

Note that the existence of the $\alpha$-parameter modifies the scalar power spectrum. For example, in the massless case $\mu=0$ and in the super-Hubble limit $y=k/\dot{a}(t)\ll1$,
\begin{eqnarray}
    \left|u_k \left(y\ll1\right)\right|^2
    =
    \left|U_k(y\ll1)\right|^2 \left( \frac{1+e^{\alpha+\alpha^{*}}-2 \mathrm{Re}(e^{\alpha})}{1-e^{\alpha+\alpha^{*}}}\right),
\end{eqnarray}
where $|U_k(y\ll1)/a|^2=H_{*}^2/2k^3$ and $a=e^{H_*t}=-1/H_*\Tilde{\eta}$. Since the scalar power spectrum is proportional to $|u_{k}/a|^2$, the mode-independent $\alpha$-parameter results in a mode-independent amplitude shift.

\subsubsection{No-boundary vs. Euclidean wormholes: Euclidean and Lorentzian initial conditions}

In the no-boundary proposal, the requirements of compactness and regularity at the South Pole uniquely pick out the Euclidean growing mode, and this leads to a Bunch-Davies mode. To allow the $\alpha$-vacuum, the other linearly independent solution, which is an Euclidean decaying mode, must be allowed. However, the Euclidean decaying mode diverges at the South Pole and breaks the regularity condition of the no-boundary instanton. This shows that the no-boundary proposal does not admit an $\alpha$-vacuum state.

On the other hand, in the Euclidean wormhole scenario, since $a(\tau=0)=a_{\mathrm{min}}\neq 0$, there would be no divergence for the two linearly independent mode functions, and they can both survive to construct the $\alpha$-vacuum mode:
\begin{eqnarray}\label{alpha-EW}
    v_{n}=N_{\alpha}\left(V_{n}+e^{\alpha^{*}}V^{*}_{n}\right)
\end{eqnarray}
in the Lorentzian domain, where $V_n$ ($V^{*}_n$) is approximately the negative (positive) frequency Bunch-Davies mode, which will correspond to a growing (decaying) mode in the Euclidean domain, and $\alpha^{*}$ is the complex conjugate of $\alpha$.

\begin{figure}[tbp]
\centering 
\includegraphics[width=.45\textwidth]{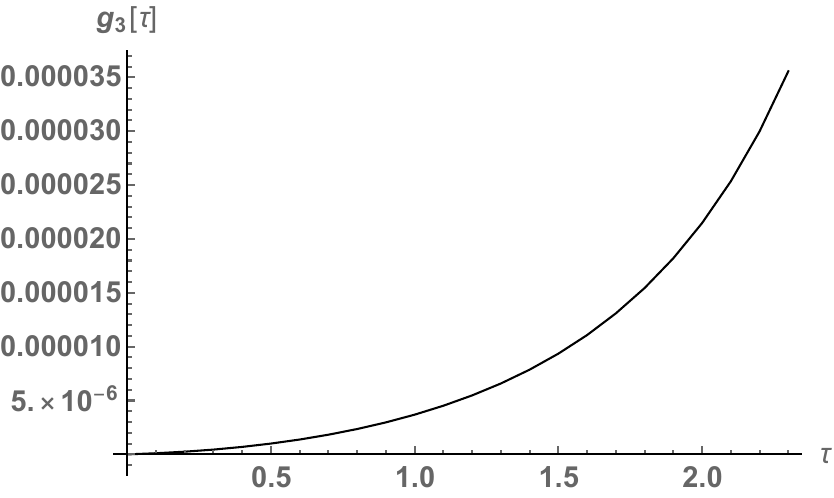}
\hfill
\includegraphics[width=.45\textwidth]{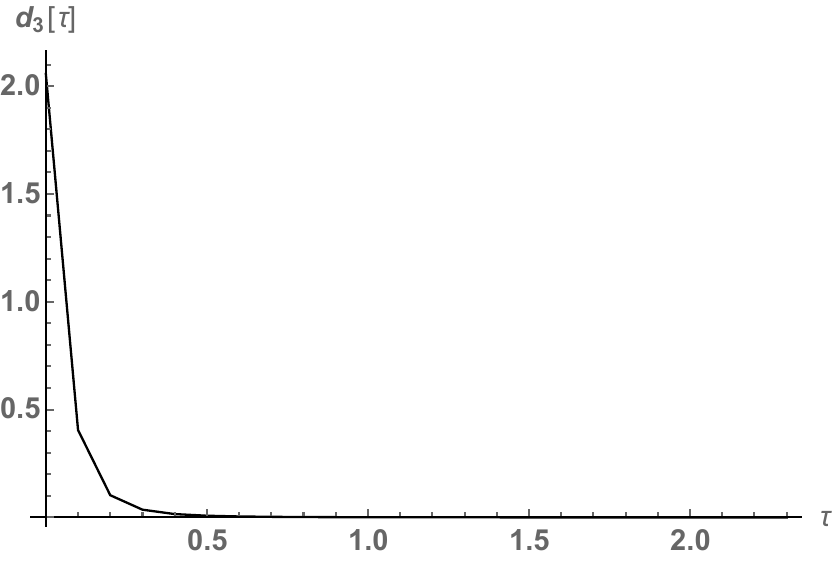}
\caption{\label{EWH-fE} Perturbation modes ($n=3$) in the Euclidean regime with $\delta_n=0$. The left is the Euclidean growing mode and the right is the Euclidean decaying mode. These two Euclidean modes can be significantly different at small $\tau$, but they can have identical values at the Wick rotation point $\tau=X$.}
\end{figure}

Let us denote the Euclidean growing (decaying) mode corresponding to $V_n/a$ $(V^{*}_n/a)$ by $g_{n}$ $(d_{n})$. Then, Lorentzian initial conditions of $v_{n}$ at $t=0$ are determined by the Euclidean modes evaluated at the Wick-rotation point $\tau=X$ by
\begin{eqnarray}
    v_{n} &=& N_{\alpha} a_{\mathrm{max}} \left( g_{n}+e^{\alpha^{*}}g^{*}_{n} \right), \\
    v'_{n} &=& iN_{\alpha} a_{\mathrm{max}}^2 \left( \dot{g}_{n}-e^{\alpha^{*}}\dot{g^{*}}_{n}\right),
\end{eqnarray}
where $'$ is the derivation with respect to $\eta \equiv 2\arctan[\tanh\big(\lambda t/2\big) ]$, $a(\tau=X) = a_{\mathrm{max}}$, and we have made use of the fact that
\begin{eqnarray}
d_n(X) &=& g^{*}_n(X), \\
\dot{d}_{n}(X) &=& -\dot{g^{*}}_{n}(X),
\end{eqnarray}
which are necessary for the analyticity. This shows that, for a wide range in the $\alpha$-parameter space, i.e., $|\mathrm{Re}(\alpha)|\geq 1$, $v_n$ can be approximately dominated by a single Bunch-Davies component, if $g_n(X)$ and $\dot{g}_n(X)$ have the same order of magnitudes as their complex conjugates, due to the exponentially hierarchy between two modes by the factor $e^{\mathrm{Re}(\alpha)}$. One can observe that if $\mathrm{Re}(g_n)$ monotonically increases forward in $\tau$, then $\mathrm{Re}(d_n)$ monotonically decreases (see Fig.~\ref{EWH-fE}). This means that, even when $|\mathrm{Re}(\alpha)|\geq 1$, it is still possible that $v_{n}$ in the Euclidean domain has different Euclidean initial conditions compared to the no-boundary proposal, depending on the value of $e^{\alpha^{*}}d_{n}(0)$.

To summarize, the Euclidean growing and decaying modes are exponentially increasing and decreasing in the Euclidean domain. In the no-boundary proposal, since the decaying mode diverges at $\tau=0$, one is forced to choose $\mathrm{Re}(\alpha)\rightarrow-\infty$, which uniquely picks out the Bunch-Davies vacuum. However, it is fair to say that Euclidean wormholes allow the adjustment of the $\alpha$-parameter to allow non-trivial $\alpha$-vacua by superposing the growing and decaying modes, since the two modes are regular.

\subsubsection{The $n$-dependent phase factor in the Euclidean regime}

Now we ask how to adjust the initial condition of scalar perturbations at $\tau = 0$. In this subsection, we discuss the detailed reason why we have chosen a specific form Eq.~\eqref{EInitialCond}.

Recall that we have included a non-trivial phase $\delta_{n}=n\pi/2$ for the Euclidean initial conditions in Eq.~\eqref{EInitialCond}. In the following, we will demonstrate that there is a non-trivial $n$-dependent phase factor in the mode function for closed universes which is consistent with the flat space limit. As a result, we need to include this phase factor $\delta_{n}$ for the Euclidean initial conditions to have the correct power spectrum for a general $\alpha$-vacuum.

In the following, we utilize a massless spectator scalar field in a pure de Sitter background to analytically show why this phase factor emerges in the closed universe setting\footnote{We are aware that the mixing of positive and negative frequency modes for the massless scalar field cannot be called an alpha vacuum since the corresponding vacuum is not invariant under the full continuous de Sitter symmetry. Nevertheless, this phase issue exists both in the massive and massless cases, and the analytic result is much easier to derive in the massless case.}. Before we enter the main discussion, we first quickly review the parallel scenario in the flat slicing de Sitter space and introduce a more useful variable that will play a crucial role in the closed universe case later. 

Firstly, the mode equation of a massless spectator scalar field in flat slicing de Sitter space is given by \cite{Baumann:2009ds, Baumann:2018muz}
\begin{equation}\label{Flat_Massless_ModeEq}
   u_{k}''(\Tilde{\eta})+\left(k^2-\frac{2}{\Tilde{\eta}^2}\right)u_{k}(\Tilde{\eta})=0,
\end{equation}
where $k$ is the comoving wave number and $\Tilde{\eta}$ is the conformal time runs from $-\infty$ to $0^{-}$. 
This differential equation admits the general solution
\begin{equation}
u_k(\Tilde{\eta})=A_{k}\frac{1}{\sqrt{2k}}\left(1-\frac{i}{k\Tilde{\eta}}\right)e^{-ik\Tilde{\eta}}+B_{k}\frac{1}{\sqrt{2k}}\left(1+\frac{i}{k\Tilde{\eta}}\right)e^{ik\Tilde{\eta}},   
\end{equation}
where the coefficients are chosen to follow the relation $|A_{k}|^2-|B_{k}|^2=1$ to have the creation and annihilation operators satisfying the required commutation relation. 

When the wavelength of a mode is much smaller compared to the Hubble radius, the curvature of the spacetime has little effect on it. In the flat slicing de Sitter space, it happens for all modes in the infinite past $\Tilde{\eta} \rightarrow -\infty$, where the mode solutions reduce to the form of the usual positive mode solution in a Minkowski spacetime
\begin{equation}\label{Minkowski_mode}
\lim_{\Tilde{\eta} \rightarrow \-\infty}u_k(\Tilde{\eta})=\frac{1}{\sqrt{2k}}e^{-ik\Tilde{\eta}}.     \end{equation}
Thus, modes in their sub-horizon limit offer the natural choice of coefficients: $A_{k}=1$ and $B_{k}=0$, which leads to the mode solution    
\begin{equation}\label{BD_flat}
u_k(\Tilde{\eta})=\frac{1}{\sqrt{2k}}\left(1-\frac{i}{k\Tilde{\eta}}\right)e^{-ik\Tilde{\eta}}.     
\end{equation}
That is, we choose only the positive frequency mode function.

Next, to make the comparison of the wavelength of a mode and the characteristic length scale of the expanding background more explicitly, we can consider the following variable
\begin{equation}\label{y_flat}
\tilde{y} \equiv \frac{k}{aH}=-k\Tilde{\eta} \sim \frac{R_{H}}{l_{phy}(k)},    
\end{equation}
where $R_{H}$ is the physical Hubble radius, and $l_{phy}(k)$ is the physical length of a mode specified by the comoving wave number $k$. Using Eq.~(\ref{y_flat}), one can rewrite Eq.~(\ref{Flat_Massless_ModeEq}) into
\begin{equation}\label{ModeEqFlat_y}
 u_{k}''(\tilde{y})+\left(1-\frac{2}{\tilde{y}^2}\right)u_{k}(\tilde{y})=0,
\end{equation}
and the positive frequency mode function Eq.~(\ref{BD_flat}) into
\begin{equation}\label{BD_flat_y}
u_{k}(\tilde{y})=
\frac{1}{\sqrt{2k}}\left(1+\frac{i}{\tilde{y}}\right)e^{i\tilde{y}}.    
\end{equation}
Notice that the condition of sub-horizon limit, where we previously determined the coefficients $A_k$ and $B_k$, becomes $\tilde{y} \gg 1$, and under this limit, Eq.~(\ref{ModeEqFlat_y}) reduces to 
\begin{equation}
u_{k}''(\tilde{y})+u_{k}(\tilde{y})=0,
\end{equation}
for which the positive frequency mode solution is
\begin{equation}\label{UV_flat}
u_{k}= \frac{1}{\sqrt{2k}}e^{i\tilde{y}}= \frac{1}{\sqrt{2k}}e^{-ik \Tilde{\eta}}.
\end{equation}
We shall refer to Eq.~\eqref{UV_flat} as the standard sub-horizon limit (for the massive case, it is called the Bunch-Davies limit) in which the mode solution is of a plane-wave form both in $\Tilde{\eta}$ and $\tilde{y}$. Meanwhile, the super-horizon limit, where we evaluate the power spectrum of the scalar field, is simply given by the limit $\tilde{y} \ll 1$. We now see how these relations play out in the closed de Sitter space.

Returning to the case of a closed universe, the equation of motion for the perturbation mode $f_{n}=v_{n}/a=u^{*}_n/a$,\footnote{Following the usual convention, we denote the positive frequency mode by $u$ and the negative frequency mode by $v$. They are simply the complex conjugate of each other. Hence, $u$ and $v^{*}$ define the same vacuum state.} Eq.~\eqref{EoM_mode}, can be cast into the following massless Lorentzian form:
\begin{equation}\label{ModeEq_closed_eta}
u''_{n}+\left[n^2-1+\frac{1}{2}\left(\cos 2\eta -3\right)\sec^2 \eta \right]u_{n}=0,   
\end{equation}
upon the standard analytic continuation and transforming the comoving time into the conventional conformal time: $\eta=2\arctan[\tanh\big(\lambda t/2\big) ]$. In this convention, $0 \leq \eta< \pi/2$. Similar to Eq.~(\ref{y_flat}), we can define the variable  
\begin{equation}\label{y_closed}
y = \frac{n}{aH}=\frac{n}{\tan\eta}\sim \frac{R_{H}}{l_{phy}(n)},
\end{equation}
and use it to rewrite Eq.~(\ref{ModeEq_closed_eta}) into 
\begin{equation}\label{ModeEq_closed_y}
\left(1+\frac{y^2}{n^2}\right)\left[(n^2+y^2) u''_{n}(y)+2y u'_{n}(y) \right]
+\left[\frac{n^2}{y^2}(y^2-2)-2\right]u_{n}=0.  
\end{equation}
This admits the positive frequency no-boundary solution \cite{Invariant vacuum}\footnote{In appendix \ref{Append}, we proved the equivalence between this solution and the massless limit of Eq.\eqref{LaflammeSol}.}:
\begin{equation}\label{BD_closed_y}
u_{n}
=
\frac{n}{\sqrt{2(n+1)n(n-1)}}\left(1+\frac{i}{y}\right)\exp \left[-in\arctan \frac{n}{y} \right], 
\end{equation}
where $n \geq 2$. Now, comparing Eq.~(\ref{ModeEqFlat_y}) with Eq.~(\ref{ModeEq_closed_y}), we see that one major difference between them is that the $k$ is removed from the differential equation Eq.~(\ref{ModeEqFlat_y}) but $n$ still exists in Eq.~(\ref{ModeEq_closed_y}). As a result, there are three special regimes in the closed de Sitter scenario instead of two as in the flat de Sitter case, i.e., $y \ll 1$ and $y \gg 1$.  Putting these three special regimes according to the time order, they are $y\gg n\gg 1$,  $n\gg y\gg1$, and $n\gg 1 \gg y$.

One can check that the mode equation Eq.~(\ref{ModeEq_closed_y}) reduces to the same form of the sub-horizon and super-horizon limits of Eq.~(\ref{ModeEqFlat_y}) in the regimes $n\gg y \gg 1$ and $n\gg 1 \gg y$ as
\begin{equation}\label{3_regimes_EOM}
\begin{split}
u_{n}''(y)+u_{n}(y)=0 \quad &when \quad n\gg y \gg 1,  \\
u_{n}''(y)-\frac{2}{y^2}u_{n}(y)=0 \quad &when \quad n\gg 1 \gg y, 
\end{split}
\end{equation}
respectively, wherein the second line we have assumed $|n^2u_n''|\gg|y u_n'|$, which can be checked by the solution whether it is consistent.
Now, the crucial point is this: does the positive frequency solution Eq.~(\ref{BD_closed_y}) also reduce to the similar forms in these two regimes as those in the flat de Sitter case?    

To answer this question, we use the following approximation in the two limits: 
\begin{equation}\label{key_approx}
\begin{split}
\arctan \frac{n}{y} 
\simeq \frac{n}{y}+\mathcal{O}\left(\frac{n^3}{y^3}\right) \quad &when \quad \frac{n}{y}\ll 1,    \\ 
\arctan \frac{n}{y} 
\simeq 
\frac{\pi}{2}-\frac{y}{n}+\mathcal{O}\left(\frac{y^3}{n^3}\right) \quad &when \quad \frac{y}{n}\ll 1.
\end{split}    
\end{equation}
Then we have Eq.~(\ref{BD_closed_y}) reducing to 
\begin{equation}\label{3_regimes_solutions}
\begin{split}
u_{n}(y) 
\sim 
\frac{1}{\sqrt{2n}}e^{-in^2/y}  \quad &when \quad y \gg n\gg 1, \\
u_{n}(y)
\sim 
\frac{1}{\sqrt{2n}}e^{-in\pi/2}e^{iy} \quad &when \quad n\gg y \gg 1,  \\
u_{n}(y)
\sim 
\frac{1}{\sqrt{2n}}e^{-in\pi/2}
\frac{i}{y}e^{iy} \sim \frac{1}{\sqrt{2n}}e^{-in\pi/2}
\frac{i}{y} \quad &when \quad n\gg 1 \gg y, 
\end{split}
\end{equation}
where the approximation $e^{iy} \sim 1$ is used to have the final form of $u_{n}(y)$ in regime $ n\gg 1 \gg y$ satisfying the super-horizon limit of the mode equation given in Eq.~(\ref{3_regimes_EOM}). We then see that the mode solution contains an extra $n$-dependent phase factor $e^{-in\pi/2}$ in the regimes we are interested in. To remove this extra factor in the regimes with $n \gg 1\gg y$, we multiply Eq.~(\ref{BD_closed_y}) by an overall phase factor $e^{i n \pi /2}$ to have
\begin{equation}\label{correct_positive_F_mode}
\tilde{u}_{n}(y)
=e^{i n \pi /2}u_{n}(y)= e^{i n \pi /2}\frac{n}{\sqrt{2(n+1)n(n-1)}}\left(1+\frac{i}{y}\right)\exp \left[-in\arctan \frac{n}{y} \right],   
\end{equation}
and conclude that $\tilde{u}_{n}(y)$ should be the correct form of the positive frequency mode solution for a massless spectator field in the closed de Sitter space. Meanwhile, we can easily see that the multiplication of $e^{i n \pi /2}$ creates an $n$-dependent phase factor to the regimes $y \gg n\gg 1$. That is,
\begin{equation}
\tilde{u}_{n}(y)
\sim 
e^{in\pi/2}\frac{1}{\sqrt{2n}}e^{-in^2/y}  \quad when \quad y \gg n\gg 1.
\end{equation}
As we mentioned earlier, this is the regime corresponding to the earliest time, and especially this extra phase in Eq.~(\ref{correct_positive_F_mode}) exists after analytic continuation to the Euclidean regime. This is the reason why we employed $\delta_n=n\pi/2$ for the initial conditions Eq.~\eqref{EInitialCond}.

\subsubsection{Agreement between analytic and numerical results}

Now, based on these initial conditions for the $\alpha$-vacuum, we report on the numerical results of the scalar power spectrum.

\paragraph{Bunch-Davies vacuum}

\begin{figure}[tbp]
\centering 
\includegraphics[width=.45\textwidth]{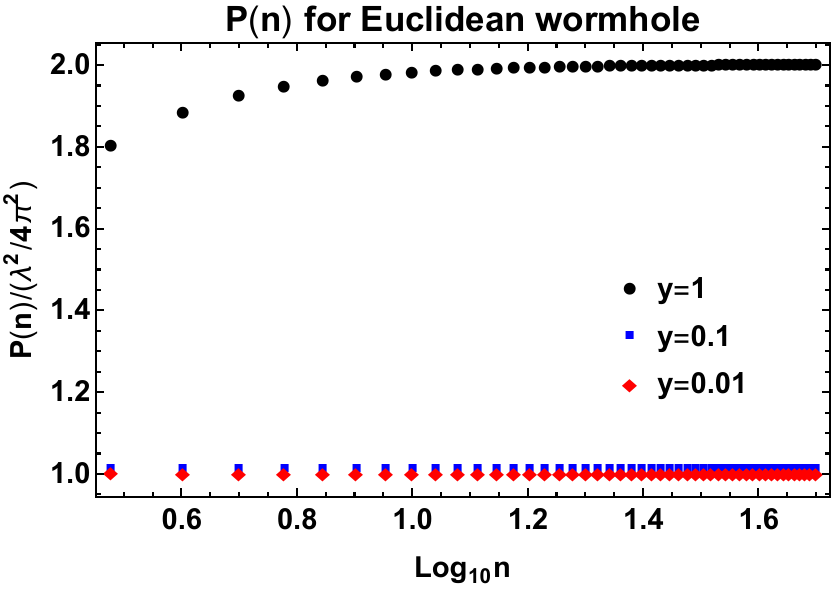}
\hfill
\includegraphics[width=.45\textwidth]{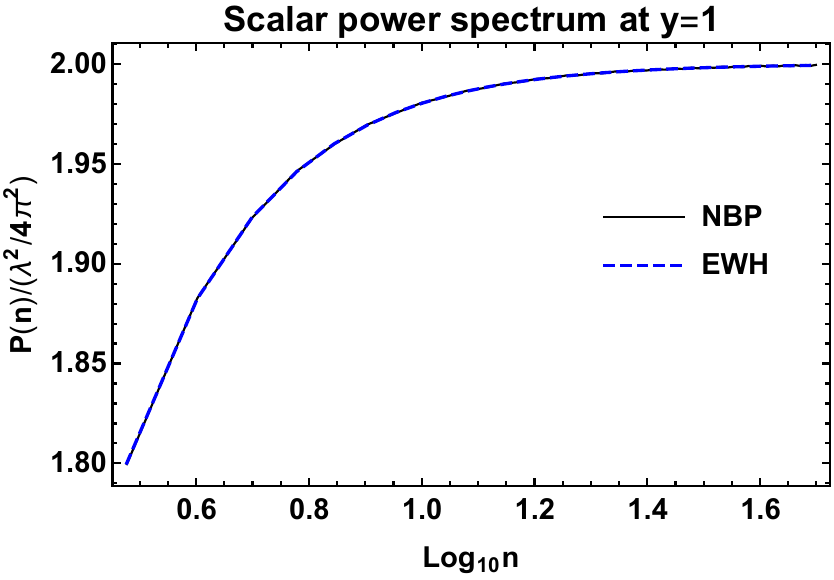}
\caption{\label{Pf} Left: Scalar power spectrum $P(n)$ in the presence of an Euclidean wormhole in a closed universe with a flat inflationary potential. In the presence of a positive spatial curvature, there is a suppression at small $n$. At large $n$, the spectrum recovers the scale-invariant behavior. Right: Comparison between the spectra for the no-boundary proposal and the Euclidean wormhole scenario with perturbation mode satisfying identical Euclidean initial conditions Eq.~\eqref{EInitialCond}.}
\end{figure}

\begin{figure}[tbp]
\centering 
\includegraphics[width=.45\textwidth]{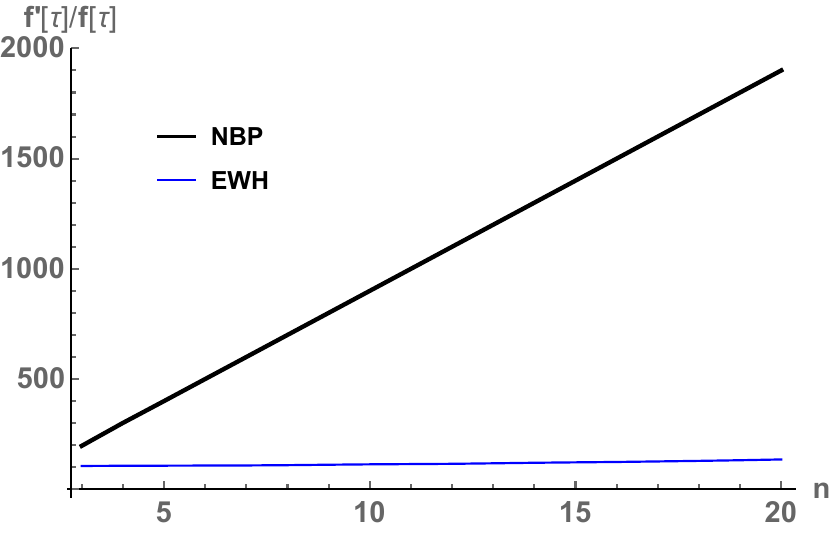}
\hfill
\includegraphics[width=.45\textwidth]{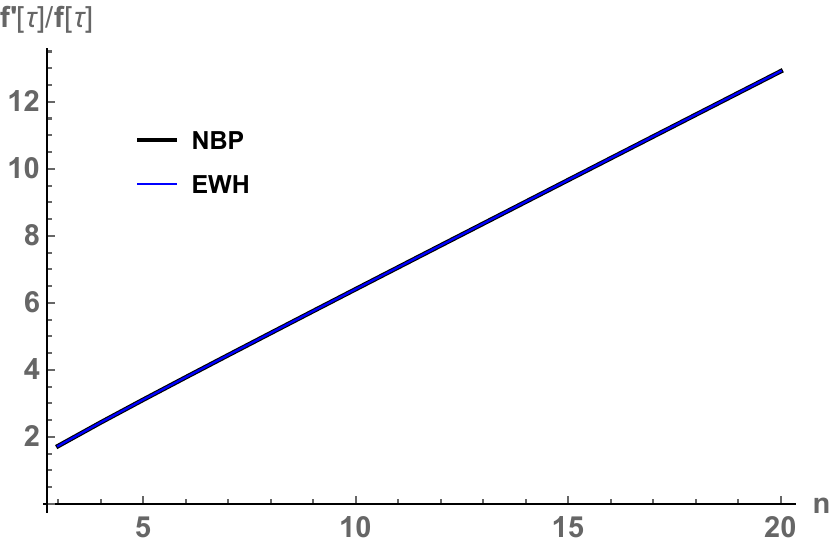}
\caption{\label{ratio} The ratio $\dot{f}_{n}/f_{n}$ at $\tau=0.01$ (left) and at the Wick-rotation point $\tau=X$ (right) in the Euclidean domain for $f_{n}$ satisfying the initial conditions Eq.~\eqref{EInitialCond} in the no-boundary proposal (black) and the Euclidean wormhole scenario (blue).}
\end{figure}

The left of Fig.~\ref{Pf} shows the scalar power spectrum $P(n)$ at different times with the Euclidean matter perturbation $f_{n}$ satisfying the initial conditions Eq.~\eqref{EInitialCond} in the presence of an Euclidean wormhole. Interestingly, the right of Fig.~\ref{Pf} shows that the power spectrum of the Euclidean wormhole is approximately identical to that of the no-boundary proposal. As indicated in Fig.~\ref{ratio}, the perturbation modes $f_{n}$ satisfying Eq.~\eqref{EInitialCond} are in general different in different Euclidean background geometries, especially at small $\tau$. However, near the Wick-rotation point, the quantity $\dot{f}_{n}/f_{n}$ (and therefore the Lorentzian initial condition) becomes approximately identical. This illustrates that a certain extreme of the results of Euclidean wormholes consistently corresponds to the result of the no-boundary proposal.

\paragraph{$\alpha$-vacuum}

In the last subsection, we considered the Euclidean wormhole power spectrum in which the Euclidean perturbation mode $f_{n}=v_{n}/a$ satisfies the initial conditions given in Eq.~\eqref{EInitialCond}. This spectrum approximates to the no-boundary (Bunch-Davies) spectrum. Hence, let us denote this solution as $f_n=g_{n}$ and $v_{n}=V_{n}$, where $g_{n}$ is the growing mode satisfying Eq.~\eqref{EInitialCond} and the equalities refer to picking out a single Bunch-Davies component among the $\alpha$-vacuum expression, Eq.~\eqref{alpha-EW}. The scalar power spectrum, in this case, is then, according to Eq.~\eqref{PowerSpectrum},
\begin{eqnarray}\label{Pn-BD-1}
    P(n)
    =\frac{n(n^2-1)}{4\pi^2\mathrm{Re}\big[-i\frac{v'_{n}}{v_{n}}\big]}
    =\frac{n(n^2-1)}{4\pi^2\mathrm{Re}\big[-i\frac{V'_{n}}{V_{n}}\big]}
    \simeq
    \frac{\lambda^2}{4\pi^2}\frac{n^2(1+y^2)}{n^2+y^2},
\end{eqnarray}
which is plotted in Fig.~\ref{Pf}.

On the other hand, according to Eq.~\eqref{alpha-EW}, the $\alpha$-vacuum mode leads to the new power spectrum:
\begin{eqnarray}\label{Pn-alpha-1}
    P(n) = S_{\alpha}\left(\frac{n(n^2-1)}{4\pi^2\mathrm{Re}\left[-i\frac{V'_{n}}{V_{n}}\right]}\right),
\end{eqnarray}
where the shifting function
\begin{eqnarray}
    S_{\alpha}=N_{\alpha}^2\left(1+e^{\alpha+\alpha^{*}}+\frac{2\mathrm{Re}[e^{\alpha}(V_{n})^2]}{|V_{n}|^2} \right)
\end{eqnarray}
is the modification to the original Bunch-Davies power spectrum when the $\alpha$-vacuum state is considered. This result is consistent with the results of the flat space limit.

\begin{figure}[H]
\centering 
\includegraphics[width=.45\textwidth]{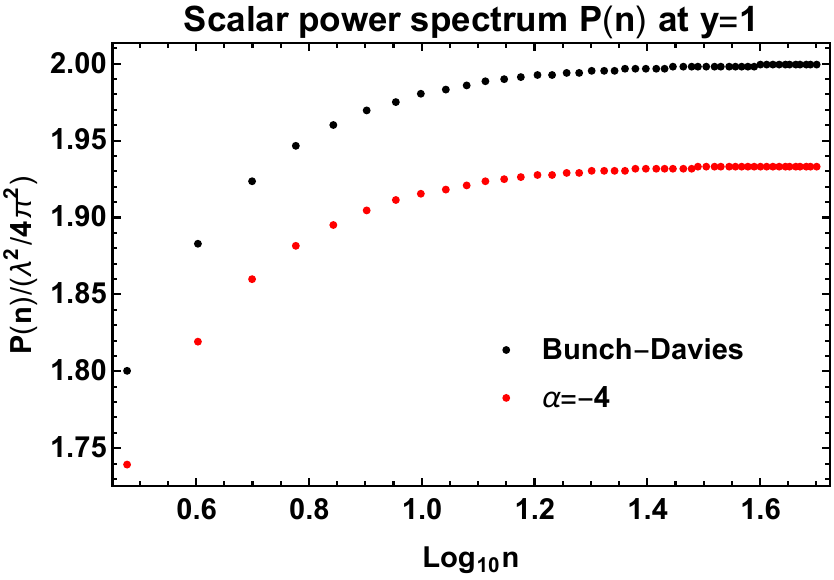}
\hfill
\includegraphics[width=.45\textwidth]{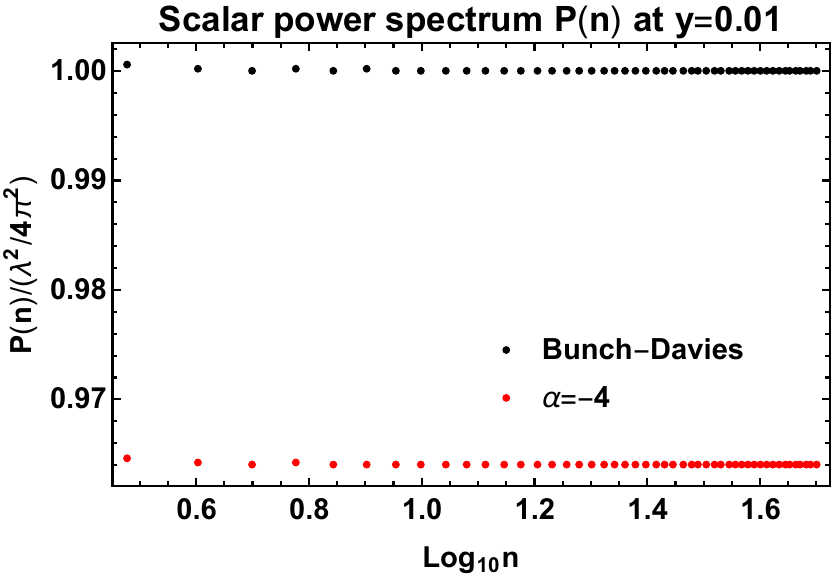}
\caption{\label{Pn} Scalar power spectrum $P(n)$ in the presence of an Euclidean wormhole in a closed universe with a sufficiently flat inflationary potential at $y=1$ (left) and $y=0.01$ (right). On the one hand, at large $n$, the spectrum becomes scale-invariant; at small $n$, the spectrum is suppressed by the curvature. On the other hand, a mode-independent $\alpha$-parameter results in an overall suppression.}
\end{figure}

Although we do not have an analytic solution for the perturbation mode in the presence of a Euclidean wormhole, it is still possible to comprehend several features in the spectrum by utilizing existing analytic no-boundary Bunch-Davies solutions, since, as previously shown in Fig.~\ref{Pf-NBP-compare}, Euclidean wormhole (in our discussion) and the no-boundary proposal have similar spectra when Bunch-Davies state is considered. Hence, multiplying Eq.~\eqref{BD_closed_y} by the phase factor $e^{in\pi/2}$ and taking its complex conjugate, $V_n=e^{-in\pi/2}u^{*}_n$, we obtain, for $\alpha\in\mathbb{R}$ and mode-independent,
\begin{eqnarray}
    \frac{\mathrm{Re}[e^{\alpha}(V_{n})^2]}{|V_{n}|^2}
    \simeq -e^{\alpha}\sin(2y)\simeq -e^{\alpha}\sin(2), \quad &when \quad n\gg y=1,
    \\
    \frac{\mathrm{Re}[e^{\alpha}(V_{n})^2]}{|V_{n}|^2}
    \simeq -e^{\alpha}\cos(2y)\simeq -e^{\alpha},\quad &when \quad n\gg 1\gg y.
\end{eqnarray}
This indicates that there will be an overall suppression in the spectrum at horizon-crossing or in the super-horizon regime, which justifies the results shown in Fig.~\ref{Pn}. This is qualitatively consistent with the conventional suppression that occurs in a flat de Sitter universe with a mode-independent $\alpha$-parameter \cite{Danielsson:2002kx,Easther:2002xe,Easther:2004vq}.

To summarize, we could obtain the scalar power spectrum of the $\alpha$-vacuum from the perturbations of Euclidean wormholes, where the results are consistent with the analytic results of the flat space limit (large $n$ limit).

\section{\label{sec:con}Conclusion}

In this paper, we followed the Euclidean path integral approach and investigated the cosmological observables under the Euclidean wormholes geometry, which is a very natural extension of the no-boundary compact instantons one.

We obtained the power spectrum formula that can be applied to the Euclidean wormholes case. To define the power spectrum, we need to provide a vacuum condition, where there is an ambiguity as to whether the vacuum is necessarily the Bunch-Davies type or the de Sitter invariant vacuum, the so-called $\alpha$-vacuum. Indeed, if we impose the compact instantons ansatz, the Euclidean vacuum uniquely points to the Bunch-Davies vacuum, whereas under a different geometry, i.e., Euclidean wormholes, a different type of vacuum, the $\alpha$-vacuum state, is allowed. This is not too surprising, because the regularity condition at the bottleneck of the wormhole is not required. This renders the quantum states of the universe a geometric origin.

We note, however, that a wide range of $\alpha$-parameter spaces would provide a vacuum state that is indistinguishable from the Bunch-Davies state. The simple reason is that one of the two linearly independent modes, i.e., the exponentially decreasing (in Euclidean time) mode, is suppressed by a factor of $e^{\mathrm{Re}(\alpha)}$; hence, in practice, only one mode can survive, and that sufficiently explains the dominance of the Bunch-Davies vacuum.

Nevertheless, the typical $\alpha$-vacuum that is distinguishable from the Bunch-Davies state remains possible, even though the parameter space is somewhat narrow. In this sense, from the perspective of the Euclidean path integral approach, the existence of the $\alpha$-vacuum is a consequence of the different geometry of the Euclidean wormholes from that of the no-boundary compact instantons.

Our ultimate goal is to embed this scenario into a realistic inflation model. We have already obtained a formula that includes curvature perturbations, where we constructed a background wormhole solution that is consistent with the inflation model. A more ambitious goal is to find out whether there exists any other observable, such as primordial gravitational waves, that can either confirm or falsify our scenario by future observations. 

Last but not least, if our universe was indeed originated from the Euclidean wormholes, then one can naturally interpret that there exist two back-to-back universes \textit{that are created from nothing}. Then, these two universes must be entangled with each other. Can the entanglement entropy between the two universes be estimated? Furthermore, if it is the case, then our universe would not be a pure state. What then would be the implications in the theoretical and observational perspectives? We leave these topics for future research efforts.

\newpage

\section*{Acknowledgment}
PC was supported by Taiwan's National Science and Technology Council (NSTC) under project number 110-2112-M-002-031, and by the Leung Center for Cosmology and Particle Astrophysics (LeCosPA), National Taiwan University. KL appreciates the Elite Doctoral Scholarship jointly provided by National Taiwan University (NTU) and Taiwan's National Science and Technology Council (NSTC). DY and WL were supported by the National Research Foundation of Korea (Grant No.: 2021R1C1C1008622, 2021R1A4A5031460).

\appendix

\section{Expectation value of matter perburbation}

In this appendix, we compute the expectation values of observables which are important to compare with cosmological observations \cite{Chen:2017aes,Yeom:2017ikw,Chen:2019mbu}.

Since the wave function for perturbation $\psi_{nlm}$ satisfies the Schrodinger equation, it can be normalized by
\begin{eqnarray}
    \int d f_{nlm} \psi^{*}_{nlm} \psi_{nlm} = 1
\end{eqnarray}
in the Lorentzian domain. Therefore, we obtain the normalization factor:
\begin{eqnarray}
\sqrt{\pi}|C_{nlm}|^2=\sqrt{\mathrm{Re} \left[ a^2 \frac{\dot{f}_{nlm}}{f_{nlm}} \right]}.
\end{eqnarray}
From this, we can derive the two-point function:
\begin{eqnarray}
    \left\langle \left| \hat{f}_{nlm} \right|^2 \right\rangle = \int d f_{nlm} f_{nlm}^2 \left|\psi_{nlm}\right|^{2} = \left. \frac{1}{2a^3 \mathrm{Re} \left[ \frac{\dot{f}_{nlm}}{f_{nlm}}\right]}\right|_{t_{\mathrm{f}}}.
    \label{fn2}
\end{eqnarray}

Now we need to sum over all modes:
\begin{eqnarray}
    \left\langle \delta\hat{\phi}^2 \right\rangle = 2\pi^2 \sum_{nlm}\sum_{n'l'm'}  \left\langle \hat{f}^{\dagger}_{nlm} \hat{f}_{n'l'm'} \right\rangle Q_{nlm}^{*} Q_{n'l'm'}.
\end{eqnarray}
By taking the average over a 3-sphere, one obtains
\begin{eqnarray}
    \left\langle \delta\hat{\phi}^2 \right\rangle_{S_{3}} \equiv \frac{\int d\Omega \left\langle \delta\hat{\phi}^2 \right\rangle}{\int d\Omega} = \sum_{nlm}\left\langle \left| \hat{f}_{nlm} \right|^2 \right\rangle = \sum_{n} n^2\left\langle \left| \hat{f}_{n} \right|^2 \right\rangle,
\end{eqnarray}
where $d\Omega = \sin^2\chi \sin\theta d\chi d\theta d\varphi$ and relying on the relations
\begin{eqnarray}
    \int d\Omega~ Q_{nlm}Q_{n'l'm'} &=& \delta_{nn'}\delta_{ll'}\delta_{mm'},
    \\
    \sum_{l m} 1 = \sum_{l=0}^{n-1}\sum_{m=-l}^{l}1 &=& \sum_{l=0}^{n-1}(2l+1)=n^2.
\end{eqnarray}
Therefore, the power spectrum of the scalar field is
\begin{eqnarray}
\left\langle \delta\hat{\phi}^2 \right\rangle_{S_3} = \sum_{n} \frac{n \mathcal{P}(n)}{n^{2} - 1}=2\pi^2\sum_{n} \frac{n P(n)}{n^{2} - 1},
\end{eqnarray}
and hence,
\begin{eqnarray}
\mathcal{P}(n) = n \left( n^{2} - 1 \right) \left\langle \left| \hat{f}_{n} \right|^2 \right\rangle = \frac{n \left( n^{2} - 1 \right)}{ 2a^{2} \mathrm{Re} \left[ -i \frac{v'_{n}}{v_{n}} \right]},
\end{eqnarray}
where $'$ denotes a differentiation with respect to $\eta$. Here, $v_n=af_n$ and $P(n)$ has the conventional normalization.

\section{Curvature perturbation and cosmological observables}

Up to now, what we have considered is the matter perturbation. To obtain the perturbations of gauge-independent observables, we consider the dimensionless and gauge-invariant comoving curvature perturbations using the flat-slicing gauge \cite{Peter:2013avv}:
\begin{eqnarray}
    \mathcal{R}(t,\chi,\theta,\varphi)=\frac{H}{\dot{\phi}} \delta\phi = \sqrt{2}\pi\sum_{nlm}\mathcal{R}_{n}(t)Q_{nlm}(\chi,\theta,\varphi).
    \label{R}
\end{eqnarray}
From this, the curvature 2-point function is then
\begin{eqnarray}
    \left\langle \hat{\mathcal{R}}^{\dagger}\hat{\mathcal{R}}  \right\rangle_{S_3} = \sum_{n}\frac{n\mathcal{P}_{\mathcal{R}}(n)}{n^2-1} = \left| \frac{H}{\dot{\phi}} \right|^2 \sum_{n}n^2 \left\langle \left|\hat{f}_{n}^2 \right| \right\rangle.
    \label{R2}
\end{eqnarray}
From this, we obtain the primordial comoving curvature power spectrum
\begin{eqnarray}
    \mathcal{P}_{\mathcal{R}}(n) =
    \left| \frac{H}{\dot{\phi}} \right|^2 \mathcal{P}(n).
    \label{PR}
\end{eqnarray}

\begin{figure}[tbp]
\centering 
\includegraphics[width=.45\textwidth]{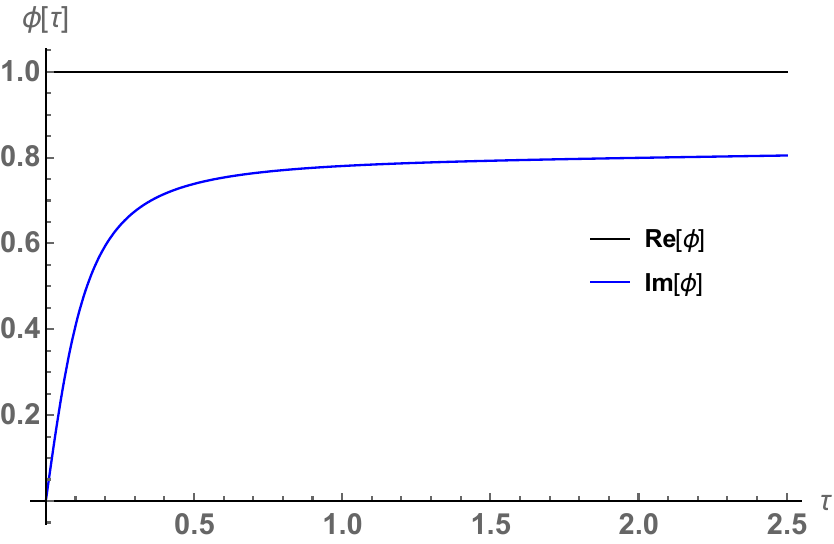}
\hfill
\includegraphics[width=.45\textwidth]{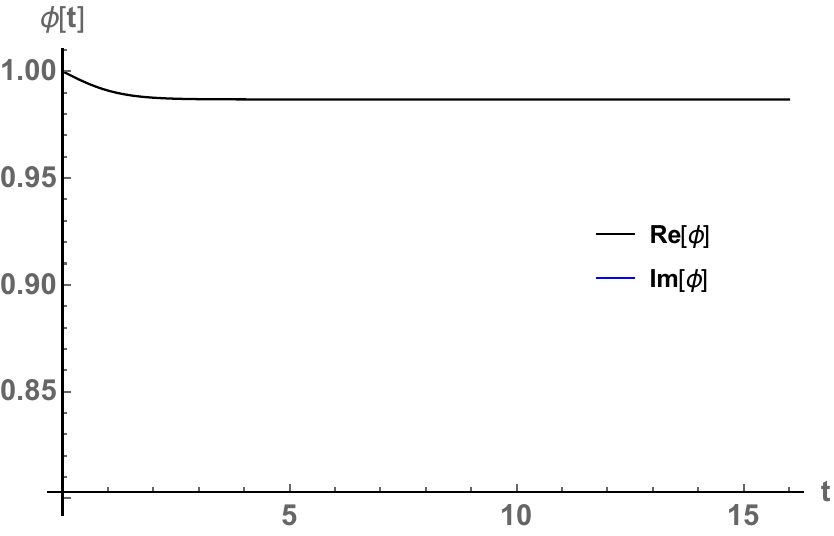}
\caption{\label{EWH-bg-phi} A demonstration of the time evolution of the inflaton field $\phi$ in the Euclidean ($\tau$) and Lorentzian ($t$) domain, where the Euclidean wormhole follows the model in \cite{Chen:2016ask,Chen:2017qeh}. The black curves are the real part and the blue curves are the imaginary part of the inflaton field $\phi$.
}
\end{figure}

In this paper, since we will only be considering a sufficiently flat inflationary potential, it does not make sense to discuss the curvature power spectrum in any more detail, as the inflation never ends and $\Dot{\phi}$ would lead to a divergence in the curvature power spectrum. Nevertheless, it is worth mentioning that the presence of an Euclidean wormhole of our type, i.e., kinetic effects of a scalar field, should induce a pre-inflationary phase that emerges in the curvature power spectrum. This may have interesting effects on small modes $n$ in the power spectrum. For example, the right of Fig.~\ref{EWH-bg-phi} shows that the Lorentzian inflation, instead of being a constant in a flat potential, has a period of non-trivial evolution in the presence of a Euclidean wormhole. The duration of this pre-inflationary phase is related to the wormhole throat $a_{\text{min}}^2\simeq A>0$ by $\Dot{\phi}=-A/a^3$. Therefore, the larger the wormhole throat, the more steep and longer the pre-inflationary phase lasts. It would be interesting to study this pre-inflationary effect (see also \cite{Chen:2017aes,Yeom:2017ikw,Chen:2019mbu}) in detail elsewhere with a more realistic inflationary potential that allows the Euclidean wormholes (e.g., \cite{Chen:2019cmw}) with which the investigation of the curvature power spectrum is meaningful.

\section{Equivalence of mode functions}\label{Append}

By using the identity for the hypergeometric function,
\begin{equation}
    {}_{2}F_{1}(n-1,n+2;n+1;\frac{1-\cos(\lambda \tau)}{2})
    =[\cos(\lambda\tau/2)]^{-2n}\big(\frac{n+\cos(\lambda\tau)}{n+1}\big),
\end{equation}
the massless limit of the regular solution Eq.~\eqref{LaflammeSol} is simplified to
\begin{equation}
    f_{n}
    =
    \frac{\lambda [\cos(\lambda \tau/2)]^{-2n}[\sin(\lambda \tau)]^{n-1}[n+\cos(\lambda\tau)]}{2^{n}\sqrt{2(n+1)n(n-1)}}.
\end{equation}
Upon analytic continuation, $\tau\rightarrow \pi/2\lambda+it$, the mode function in the Lorentzian domain is found out to be
\begin{equation}
    f_{n}
    =
    \frac{\lambda[\cosh(\lambda t)]^{n-1}[1-i\sinh(\lambda t)]^{-n}[n-i\sinh(\lambda t)]}{\sqrt{2(n+1)n(n-1)}},
\end{equation}
where we have normalized the mode function such that it has the Wronskian: $f_{n}\Dot{f}_{n}^{*}-f_{n}^{*}\Dot{f}_{n}=-i/(\lambda^{-1}\cosh(\lambda t))^{3}$.

In terms of $y=n/aH$, $t=\lambda^{-1}\text{arcsinh}(n/y)$, the mode function becomes
\begin{equation}
    v_{n}
    =
    af_{n}
    =
    \frac{n}{\sqrt{2(n+1)n(n-1)}}
\big(1-\frac{i}{y}\big)
\big(1+\frac{n^2}{y^2}\big)^{n/2}
\big(\frac{iy}{n}\big)^{n}
\big(1+\frac{iy}{n}\big)^{-n}.
\end{equation}
To simplify this, by defining
\begin{equation}
    e^{inX}
    \equiv
    \big(1+\frac{n^2}{y^2}\big)^{n/2}
\big(\frac{iy}{n}\big)^{n}
\big(1+\frac{iy}{n}\big)^{-n},
\end{equation}
taking logarithm on both sides, and using the identity: $\arctan(x)=\frac{i}{2}\log\big(\frac{1-ix}{1+ix} \big)$, one finds $X=\arctan(n/y)$. Therefore, one proves that
\begin{equation}
     v_{n}
     =
     af_{n}
    =
    \frac{n}{\sqrt{2(n+1)n(n-1)}}
\big(1-\frac{i}{y}\big)e^{in\arctan(n/y)}
\end{equation}
is nothing but the complex conjugate of Eq.~\eqref{BD_closed_y}.

\end{document}